\documentclass[preprint,showpacs]{revtex4}
\input epsf
\begin{document}
\title{Experimental investigation of optical atom traps with a frequency jump}
\author{P Ahmadi, G Behinaein, B P Timmons and G S Summy}
\address{Department of Physics, Oklahoma State University,
Stillwater, Oklahoma 74078-3072}

%\ead{peyman.ahmadi@okstate.edu}

\begin{abstract}
We study the evolution of a trapped atomic cloud subject to a
trapping frequency jump for two cases: stationary and moving
center of mass. In the first case, the frequency jump initiates
oscillations in the cloud's momentum and size. At certain times we
find the temperature is significantly reduced. When the
oscillation amplitude becomes large enough, local density
increases induced by the anharmonicity of the trapping potential
are observed. In the second case, the oscillations are coupled to
the center of mass motion through the anharmonicity of the
potential. This induces oscillations with even larger amplitudes,
enhancing the temperature reduction effects and leading to
nonisotropic expansion rates while expanding freely.
\end{abstract}


%\submitto{\jpb}

\maketitle

\section{Introduction}
A trapped atomic cloud subjected to a sudden change in its
trapping potential can exhibit a wide range of behavior with many
possible applications. It can be used as a model for studding
fundamental phenomenon ranging from superfluidity to the
generation of non-classical states of a matter-wavepacket. For
instance, a Bose-Einstein condensate can be reversibly created by
suddenly changing the trapping potential \cite{Stamper-Kurn1};
sinusoidally moving a magnetic trap center can be used to observe
the excitation modes of a Bose-Einstein condensate
\cite{Stamper-Kurn2}; the modulation of optical traps can be used
for coherent control of the center of mass wavepacket \cite{rudy};
and the breathing mode oscillation of atomic clouds in 1D and 3D
optical lattices can be realized by suddenly changing the lattice
depth \cite{Rathel}. The later has also captured a great deal of
theoretical interest because of the possibility of creating
squeezed states of a matter wavepacket \cite{Janszky1, Janszky2,
Aliga, Lo, Agarwal, Janszky3, Leibscher}.

The oscillation of an atomic cloud after a sudden jump in the
trapping frequency has also been proposed as a tool for optical
cooling and hence controlling the onset of Bose-Einstein
condensation \cite{Balatov, Baltove2}. This cooling method can be
categorized among the techniques that use geometrical manipulation
of an external potential to reduce the temperature of atoms or
molecules. For example, Flories \textit{et al.} \cite{Flories}
used a spatially varying electric field to longitudinally cool a
molecular beam to 250 nK, while Ketterle and coworkers achieved
pico-Kelvin temperatures by adiabatically decompressing a
gravito-magnetic trap \cite{Ketterle-coldatoms}. Exposing a freely
expanding cloud of cold atoms to a pulsed potential has also been
considered as a method for temperature reduction \cite{Hubert}.

In this paper we use optical traps to experimentally realize a
potential with a sudden frequency jump so as to explore the
cooling proposals of Ref. \cite{Balatov}. These traps have become
versatile tools for atom optics research, with the observation of
all-optical Bose Einstein condensation an example of their
dramatic potential \cite{ChapmanBEC}. Early experiments on
intensity modulation of the laser beam used to form an optical
trap \cite{Gorlitz,Rathel,Rudy,Monroe} were used to demonstrate
the parametric excitation of an atomic cloud. In this paper we
will discuss experiments with these traps, comparing the
observations with theory, and presenting possible future
applications. The format of this paper is as follows. In Section
II the experimental configuration is reviewed. In Section III we
first lay out the theory of non-adiabatic cooling of an atomic
cloud, then follow by presenting experimental data for the case of
a sudden frequency jump with a stationary center of mass. This
discussion is then repeated in Section IV for the case of a moving
center of mass.

\section{Experimental configuration}

Our experimental apparatus has been described previously
\cite{peymanpra, peyman} so only a brief description is given
here. We create a potential for Rb87 atoms using their interaction
with 10.6 $\mu$m light from a $\rm CO_2$ laser. The $\rm CO_2$
laser radiation frequency is far enough below that of the
resonances of the atoms that its electric field can be considered
quasi-static. This induces a dipole moment proportional to the
local electric field and a lowering of the atomic ground state
energy given by, $U=-{1 \over 2} \alpha_{g} |E|^{2}$, where
$\alpha_{g}$ is the ground state static polarizability and
$|E|^{2}$ is the time averaged square of the laser light's
electric field.

Two $\rm CO_{2}$ beams were directed into a vacuum chamber in a
geometry such that they propagated orthogonally to each other. One
of the beams propagated in the vertical direction ($x$ direction)
and the other in the horizontal direction ($z$ direction).
Crucially the position where the $x$ beam intersected the $z$ beam
could be adjusted. That is, the foci of the two beams did not
necessarily coincide. The light for these beams originated from a
50 Watt RF excited $\rm CO_{2}$ laser whose total power was
controlled by passing the output light through an acousto-optic
modulator (AOM). The first order beam of the modulator was then
directed into another AOM which was used as a beamsplitter whose
ratio could be changed by varying the input RF power. This ratio
was used to control the time dependent potential. In order to
fulfill the conditions necessary for this experiment, the $\rm
CO_{2}$ beams were aligned such that the $x$ beam was the first
order of the second AOM and the $z$ beam the zeroth order of the
same AOM. The data were taken by destructively imaging the cloud
using a resonant probe laser which passed through the atom cloud
and was then incident on a CCD camera. A gaussian fit to the
optical density data determined the spatial extent of the atomic
cloud.

\section{Stationary Center of Mass}

We begin by theoretically considering the situation where the
center of mass of the atomic cloud remains stationary after a
sudden change in the trapping potential. We take a sample of atoms
in thermal equilibrium inside a harmonic potential of frequency
$\omega_0$ which is non-adiabatically changed to $\omega_1$ at
$t=0$. Neglecting the effect of collisions and following Balatov
\textit{et al.} \cite{Balatov}, the momentum and position
probability distribution widths are given by,
\begin{eqnarray}
\sigma_p^2(t)&=&{1\over 2}\sigma_p^2(0)[1+({\omega_1\over
\omega_0})^2+(1-({\omega_1\over \omega_0})^2)\cos(2\omega_1 t)]\\
\sigma_z^2(t)&=&{1\over 2}\sigma_z^2(0)[1+({\omega_0\over
\omega_1})^2+(1-({\omega_0\over \omega_1})^2)\cos(2\omega_1 t)]
\label{eq:dispersion}
\end{eqnarray}
where at $t={\pi \over 2\omega_1}$ they reduce to,
\begin{eqnarray}
\sigma_p^2({\pi\over 2\omega_1})&=&({\omega_1\over \omega_0})^2
\sigma_p^2(0) \label{eq:momentum} \\ \sigma_z^2({\pi \over
2\omega_1})&=&({\omega_0\over \omega_1})^2 \sigma_z^2(0).
\label{eq:zero}
\end{eqnarray}
For $\omega_1 \ll \omega_0$, a narrow momentum distribution is
produced corresponding to a lower effective temperature. In the case
of a single Gaussian laser beam propagating in the $z$ direction the
average electric field can be expressed as,
\begin{equation}
|E(x,y,z)|^{2} = E_{0}^{2} \frac{w_{0}^{2}}{w(z)^{2}} {\rm exp}{-
2(x^{2} + y^{2}) \over w(z)^2}
\end{equation}
where $w(z) = w_0 \left( 1+({z \over z_R})^2 \right)^\frac12$ and
$E_0$ is the electric field amplitude. Here $w_0$ is the beam
waist at the focus, and $z_R$ is the Rayleigh length. By carrying
out a series expansion around $x=y=z=0$ and discarding terms of
third order and higher, a harmonic approximation of the potential
is obtained,
\begin{equation}
U(x,y,z) = U_{0}\left({2(x^{2} + y^{2}) \over w_0^2} + {z^2 \over
z_R^2}\right) \label{eq:potential}
\end{equation}
where $U_{0}=-{1 \over 2} \alpha_{g} E_{0}^{2}$. The oscillation
frequencies in the three directions are $\omega_x^2 = \omega_y^2
\simeq 4 U_0/m w_0^2, \omega_z^2 \simeq 2 U_0 / m z_R^2$. When the
total laser power is shared with a second beam propagating in the
$x$ direction which crosses the first beam at its focus, these
frequencies change to $\omega_y^2 \simeq 4 U_0/m w_0^2, \omega_x^2
= \omega_z^2 \simeq 2 U_0/m w_0^2 $  at the intersection point.
This assumes that the beams have identical properties and $z_R \gg
w_0$. According to these results, an abrupt change from a two beam
to a one beam geometry will produce a significant frequency change
along the $z$ direction. The ratio of one and two beam frequencies
is,
\begin{equation}
{\omega_{\rm{1 beam}}\over \omega_{\rm{2 beam}}} = {w_0 \over z_R}
= {\lambda \over \pi w_0}
\label{ratio2}
\end{equation}
which is independent of the total laser power and is determined by
the beam waist at the focus. Since the temperature of the atomic
cloud is proportional to the momentum distribution squared,
combining Eqs. (\ref{eq:momentum}) and (\ref{ratio2}) gives
\begin{equation}
{T_{\rm{final}} \over T_{\rm{initial}}} = \left( {\omega_1 \over
\omega_0}\right)^2= \left({\lambda \over \pi w_0}\right)^2.
\label{temp}
\end{equation}

To explore this behavior experimentally we aligned the $\rm CO_2$
beams such that their foci overlapped. Atoms were loaded into the
potential minimum created by the intensity maximum located at the
intersection of the beams. The power in the $x$ beam was
transferred to the $z$ beam by switching the RF power on the
second AOM to zero. This abrupt transfer of power mostly produced
a change in the effective frequency in the $z$ direction and
initiated the oscillation of the trapped cloud. If the shift of
power to the $z$ beam happens immediately after loading, the FORT
will not produce the desired oscillation since the atomic density
is approximately $10^{14}$ atoms $\rm {cm^{-3}}$ and cold binary
scattering dramatically affects the evolution of the cloud.
Therefore the collisionless approximation of the previous section
can not be used. To move into the collisionless regime it is
necessary to decrease the atomic density by reducing the number of
atoms before the power transfer. This is accomplished by reducing
the total power in the $\rm{CO_2}$ beams in order to carry out
forced evaporation of hot atoms. This reduces both the density of
atoms in the trap and their temperature. This process is carried
out so that the atomic cloud which remains in the FORT has
densities of around $10^{11}$ atoms $\rm cm^{-3}$. This guarantees
that the mean free path of the individual particles is bigger than
the cloud size so that the evolution of the sample can be
considered in the collisionless regime.

\begin{figure}
\begin{center} \mbox{\epsfxsize 3.2in \epsfysize 3.5in\epsfbox{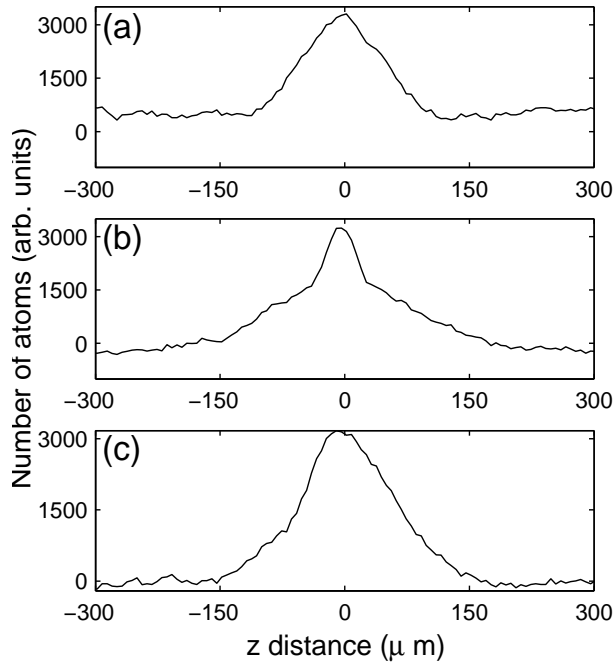}} \end{center}
\caption {The accumulated intensity along the $x$-axis as a
function of $z$ distance. Two beams from a $\rm CO_2$ laser
crossed each other at right angles with overlapping foci and 5
Watts power. (a), (b) and (c) were obtained 2, 5 and 10 ms after
extinguishing the $x$ beam. A denser component is obvious at the
center of the cloud in (b).} \label{twocomponent}
\end{figure}

\begin{figure}
\begin{center} \mbox{\epsfxsize 3.0in\epsfbox{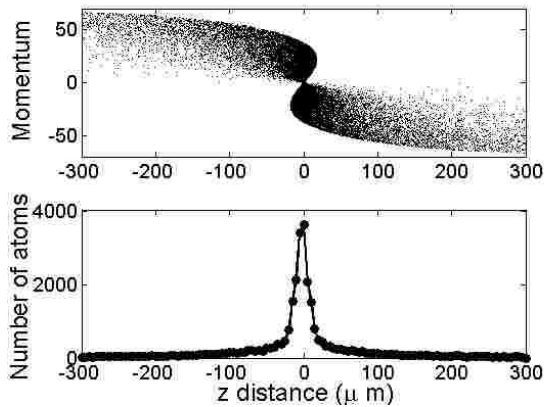}} \end{center}
\caption {(a) Calculated $z$ direction phase space distribution of
the atoms at the moment where the two component structure appears.
The atoms where initially distributed uniformly throughout the
phase space and physically located at the cross over of two beams
with overlapping foci. The horizontal axis is the $z$-direction
spread of the particles in units of the beam waist and the
vertical axis is the $z$-momentum in units of the photon recoil
from a 780 nm photon. (b) The corresponding histogram of the
particles' distribution in the $z$ direction.}
\label{twocomponentsim}
\end{figure}

After evaporative cooling the total power was transferred to the
$z$ beam and the cloud size started to oscillate in the $z$
direction. However, the time evolution of the cloud size did not
obey the simple periodic form predicted by Eq.
(\ref{eq:dispersion}). Our observations show that a domain with
high atomic density appears at the center of the trap periodically
during the cloud's evolution . This structure is reminiscent of
the two component structure which appears at the onset of a
degenerate Bose gas. Three cross sections of the cloud along the
$z$-axis are shown in Fig.\ref{twocomponent}.
Figure\ref{twocomponent} (a), (b) and (c) were taken 2, 5 and 10
ms after switching off the $x$ beam respectively. The total power
in the $\rm {CO_2}$ beams was 5 Watts. A high density region at
the center of Fig. \ref{twocomponent}(b) is clearly visible. It
disappears later in the evolution as seen in Fig.
\ref{twocomponent}(c). Creation of these high density regions is a
direct result of the strong anharmonic potential and is predicted
in \cite{Baltove2}. A Monte Carlo simulation of the anharmonic
system can reproduce this behavior for the cloud. Figure
\ref{twocomponentsim} presents the result of such a collisionless
calculation at the moment where the two component structure
appears. The simulation was carried out for $3\times 10^4$ atoms
initially at thermal equilibrium with a $25 \mu \rm m$ initial
cloud size. It is worth noting that the high density regions never
appeared when harmonic potentials were simulated.

\begin{figure}
\begin{center} \mbox{\epsfxsize 3.0in\epsfbox{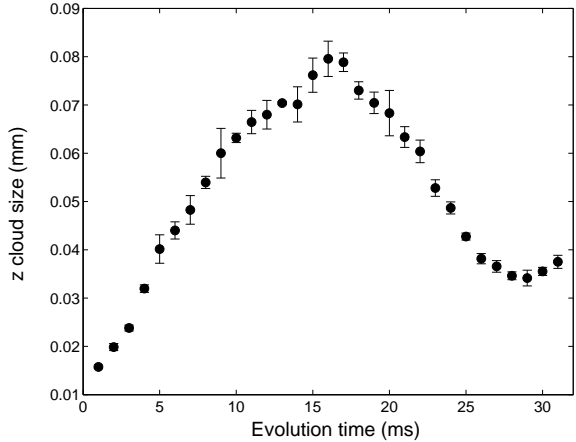}} \end{center}
\caption {Evolution of the $z$ cloud size inside the $z$ beam for
a stationary center of mass. Two beams from a $\rm CO_2$ laser
crossed each other at right angles with overlapped foci. The power
in the $x$ beam was abruptly transferred to the $z$ beam at t=0 to
initiate the cloud's oscillation.} \label{symmetriccloud}
\end{figure}

The high density region makes the observed optical density
profiles complex enough that it is impossible to fit a Gaussian
function to the data and infer the width of the cloud. Therefore
to observe the oscillation of the cloud size, predicted by Eq.
(\ref{eq:dispersion}) we have reduced the total power in the $\rm
{CO_2}$ beams to 0.5 Watts. In this regime the cloud temperature
is low enough (below 2 $\mu$K) that the oscillation amplitude does
not move the atoms away from the region where the harmonic
approximation of the potential is valid. Furthermore, as mentioned
above, the evaporative cooling also reduces the atomic density to
$\approx 10^{11}$ atom $\rm cm^{-3}$, suppressing the two-body
collision rate. The time dependency of the cloud size for this
condition is given in Fig. \ref{symmetriccloud}. This shows that
the cloud does not collapse to its original extent during its
evolution in the trap. We expect that this is the result of a
small number of collisions. According to this data the size of the
cloud in the $z$ direction expands to about 6 times its initial
value. From Eq. (\ref{eq:momentum}) this implies a factor of 6
reduction in the width of the momentum distribution or a factor of
36 in the effective $z$-direction temperature. This number in
combination with Eq. (\ref{temp}) determines the beam waist at
$\approx 20 \mu \rm m$, in good agreement with the theoretically
calculated value of $\approx 23 \mu \rm m$.

\section{Moving Center of Mass}

\begin{figure}
\begin{center} \mbox{\epsfxsize 3.0in \epsfysize 3.in \epsfbox{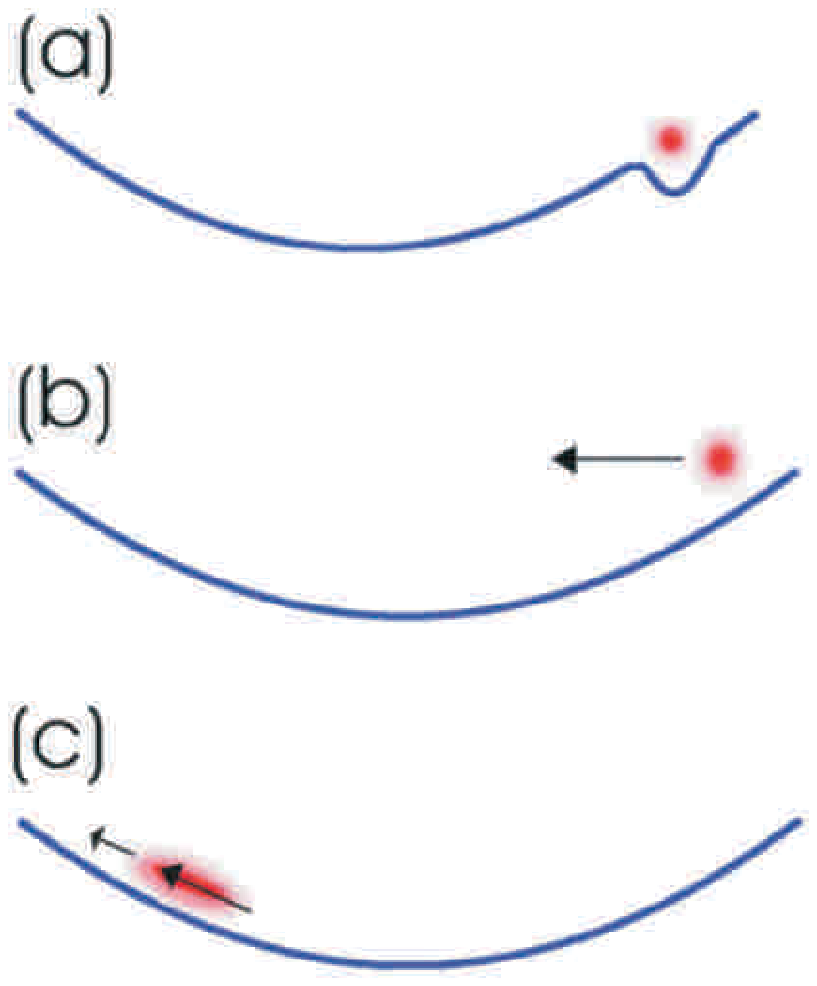}} \end{center}
\caption {Schematic of the dipole potential (a) before  and (b)
after transferring the power of the $x$ beam to the $z$ beam. In
(b) the center of mass of the atomic cloud oscillates around the
Gaussian focus of the remaining light. (c) cloud size after
evolving inside the $z$ beam for approximately half a period. The
difference in velocities at the two ends of the cloud gives rise
to a collapse after the potential is switched off for a TOF
experiment} \label{trapgeometry}
\end{figure}

In contrast to most other work with FORTs, we also studied
geometries where the center of mass moves as a result of a change
in the potential. This was achieved by offsetting the foci of the
$\rm CO_2$ beams from each other. In this configuration, when one
of the beams was switched off, the atoms moved towards the new
potential minimum located at the focus of the remaining beam. For
a harmonic trap the center of mass and internal dynamics of the
atomic cloud are decoupled from one another. However, for an
anharmonic trap the periodic motion of the center of mass of the
atoms couples to the internal dynamics. Figure \ref{trapgeometry}
schematically displays the trapping potential before and after
power is transferred to the $z$ beam.

\begin{figure}
\begin{center} \mbox{\epsfxsize 3.5in\epsfbox{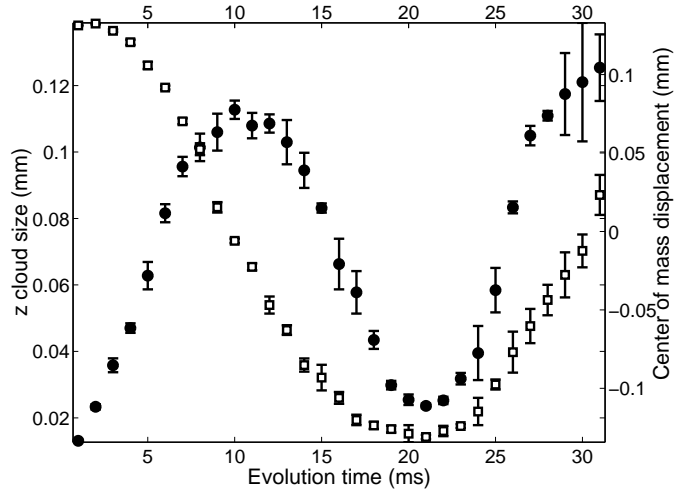}} \end{center}
\caption {The cloud size (closed circles) and center of mass
position (open squares) in the $z$ direction while evolving inside
the $z$ beam. Initially two beams from a $\rm CO_2$ crossed each
other at right angles with their foci displaced by $\approx$ 140
$\mu$m. The power in the $x$ beam was abruptly transferred to the
$z$ beam at t=0.} \label{cloudevolution}
\end{figure}
\begin{figure}
\begin{center} \mbox{\epsfxsize 3.0in \epsfysize 3.5in \epsfbox{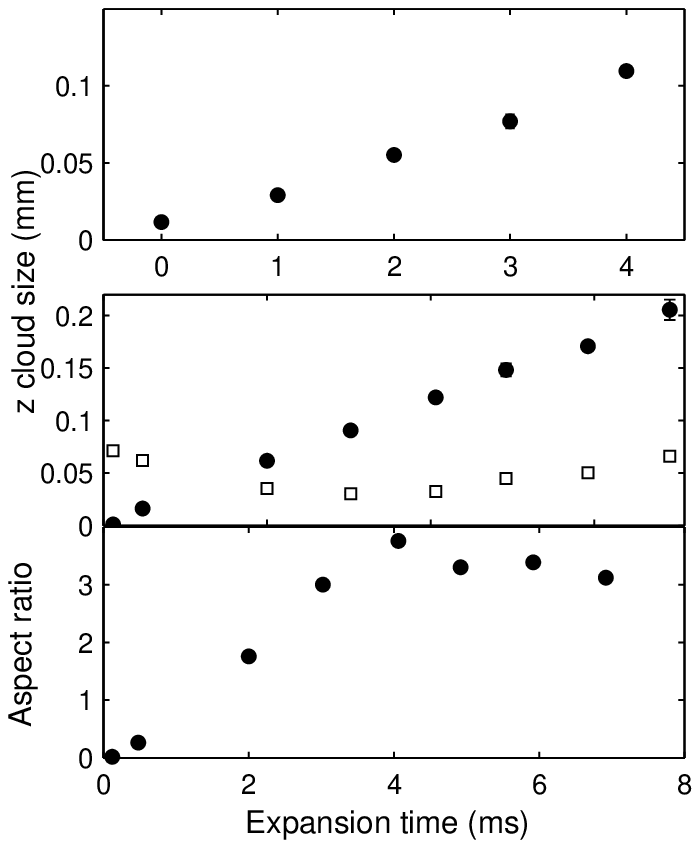}} \end{center}
\caption {Experimental data showing the cloud size. The power was
5 Watts per $\rm CO_2$ beam. The circular and square symbols
represent the cloud size in the $x$ and $z$ directions
respectively. (a) shows the results after various expansion times
for the case where the FORT has been directly released from the
crossed beams. Note that the circle and square symbols overlap
each other in this case. In (b) the $x$ and $z$ cloud sizes are
shown for the condition where the power from the $x$ beam was
switched to the $z$ beam 15 ms before releasing the FORT. In (c)
the ratio of the $x$ to $z$ cloud sizes of (b) is presented.}
\label{onewattdata}
\end{figure}
The observed $z$ cloud size as a function of time for one period
is presented in Fig. \ref{cloudevolution}. As occurred in the
previous section, the cloud size in this direction never shrinks
to its initial value. Figure \ref{cloudevolution} also shows that
the cloud undergoes a considerable increase in width by the time
it reaches the focus of the $z$-beam at t $\approx 10 ms$. It can
be seen that the cloud size increases by as much as 10 times its
initial value. This implies a compression in momentum spread by a
similar factor and thus we expect a two order of magnitude
reduction of the effective temperature in the $z$-direction. This
is significantly more than was possible in the stationary center
of mass experiments.

We also performed experiments in which the potential was switched
off completely after the atoms had evolved for various times
inside the $z$-beam.There are several features of the cloud's
evolution after release from the $z$-beam that are worthy of note.
Firstly, it is possible to tune the center of mass velocity in the
$z$-direction with high precision. This can be accomplished by
changing the time span that the cloud evolves inside the $z$-beam
before it is switched off. Such a property could be useful for
quantum reflection experiments where precise control of the impact
velocity is required \cite{Japha}. A second aspect of the free
evolution is the observation of a collapse of the cloud in the
$z$-direction during the first few milliseconds after the atoms
are released. To observe this collapse the cloud is first allowed
to evolve inside the $z$-beam for nearly half a period of the
center of mass motion. The collapse can be explained by the
velocity difference between the atoms on either side of the cloud
imprinted by the gradient of the trapping potential. This can be
seen in Fig. \ref{trapgeometry}(c). The exact time at which to
switch the $z$-beam off to observe the maximum collapse is
determined experimentally. This time is 16 ms for the experiment
shown in Fig. \ref{cloudevolution}. The resulting collapse or
focusing can be used to transfer energy from the $z$ to the $x$
and $y$ directions if the cloud has high enough densities such
that the free expansion starts when the cloud is in the
collisional regime. Under this condition, since at the release
time the trap has become elongated in the $z$ direction, the atoms
moving along the $z$ direction experience a larger number of
collisions compared to the $x$ and $y$ directions and part of
their $z$ kinetic energy transfers into $x$ and $y$ while
expanding \cite{wu}. To explore this possibility, unlike in the
previous section, the evaporative cooling is carried out such that
the cloud had a density of $\approx 10^{14}$ atoms $\rm cm^{-3}$
(this corresponds to a mean free path of $9 \mu$m for our system).
The velocity distribution should remain constant with time after
the expansion makes the mean free path larger than the cloud size
and the atoms reach the collisionless regime. Therefore to
determine the temperature the asymptotic expansion rate of the
cloud data must be used \cite{shvarchuk} after it has reached the
collisionless regime .

\begin{figure}
\begin{center} \mbox{\epsfxsize 3.0in \epsfbox{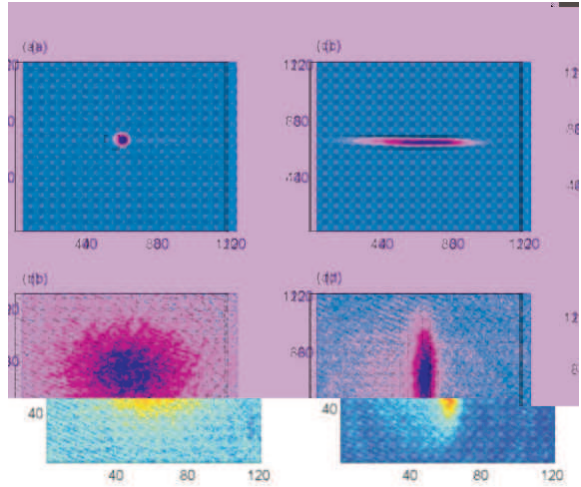}} \end{center}
\caption {Images of the atomic cloud. (a) shows cloud at the
beginning of the time of flight released from two crossed beams. In
(b) the same cloud after 4 ms free expansion. (c) released from the
crossed beam FORT 6 ms after turning off the $x$ beam and (d) is the
same cloud imaged 4 ms after the $x$ beam has been extinguished.
Each pixel in the images is 6 $\times$ 6 $\mu$m.}
\label{time-of-flight}
\end{figure}

\begin{figure}
\begin{center} \mbox{\epsfxsize 3.0in\epsfbox{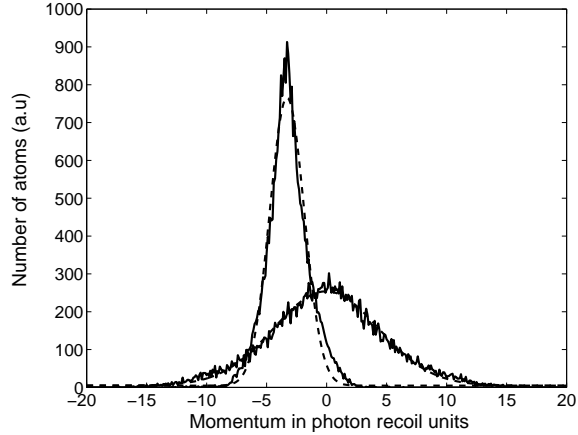}} \end{center}
\caption {Simulated momentum distributions in $z$ and $x$
directions after 5 ms free expansion. Broken lines are the
gaussian fit to the data. The narrow and broad curves are the $z$
and $x$ distributions respectively.} \label{simulatedevolution}
\end{figure}

For comparison we have conducted another series of experiments in
which the power in the $\rm{CO_2}$ laser was switched off without
transferring total power to the $z$ beam. Figure
\ref{onewattdata}(a) shows the size of the atomic cloud as a
function of time after release from a crossed beam trap with 5
Watts of power. As can be seen, the expansion of the atoms is
isotropic ($x$ and $z$ data overlapp) and the corresponding
temperature is about 3.7 (2) $\mu$K. In Fig. \ref{onewattdata}(b)
the cloud size as a function of time is given for an experiment in
which the atoms were released from the crossed beam trap (off-set
foci) and then allowed to evolve for 6 ms inside the $z$ beam
before all potentials were switched off and the free expansion
began. The asymptotic velocity for the elongated trap extracted
from data given in Fig. \ref{onewattdata}(b) gives $T_z =
0.43(2)~\mu$K and $T_x= 4.4 (2)~ \mu$K. The higher temperature in
the $x$ direction is caused by the focusing of the cloud and the
extra collisions atoms undergo in the $z$ direction because of the
compressed cloud. The ratio is $T_x/T_z = 10.2$ which is quite
close to the square of the mean aspect ratio = 3.3 for the
collisionless expansion. Figure  \ref{onewattdata}(c) shows the
ratio of the $x$ to $z$ size of the cloud at different times.
Figure  \ref{time-of-flight} shows the atomic cloud's image for 4
different times during the free expansion. Figure
\ref{time-of-flight}(a) shows the cloud at the crossed beam
without expansion and Fig. \ref{time-of-flight}(b) is the same
cloud after 4 ms free expansion. Fig. \ref{time-of-flight}(c) is
taken 6 ms after turning the vertical beam off and Fig.
\ref{time-of-flight}(d) is the same cloud after 4 ms of free
expansion.
The accurate simulation of the individual atomic trajectories for
the free expansion is very time consuming because of the two-body
cold collisions. In order to overcome this problem we employ the
direct simulation Monte Carlo (DSMC) method which was initially
developed by Bird \cite{Bird} to simulate molecular gas dynamics.
This method has been used for direct simulation of evaporative
cooling \cite{wu1} and free expansion of the cloud of atoms
\cite{wu}. Using this method the behavior of a cloud of atoms with
2 $\mu$K temperature initially displaced $0.5 z_R$ from the
potential minimum created by a gaussian beam with 25 $\mu$m beam
waist was simulated. Figure  \ref{simulatedevolution} shows the
simulated momentum distribution in the $z$ (narrow distribution)
and $x$ (broad distribution) directions after 5 ms free expansion.
The initial density was taken $1.8 \times 10^{14}$ atom $\rm
cm^{-3}$. As one can see the axial distribution deviates little
from a perfect gaussian. This has been reported previously for an
expanding cloud \cite{wu}.

\section{Conclusion}
In this paper we have studied the effect of geometrical changes of
the trap on the atomic cloud evolution. We experimented with two
configurations of crossed laser beams. In the first set of
experiments, the foci of two orthogonal beams were positioned to
conincide. The foci of the beams were subsequently offset for the
second set of experiments. In each case, the behavior of the cloud
upon abruptly switching off the $x$ beam was studied. In the
overlapped foci experiment we observed high atomic density
domains, due to the strong anharmonic trapping potential. To reach
the collisionless regime the trap density was reduced by lowering
the power in the beams. We observed oscillations in the axial $z$
direction with an amplitude up to a factor of 6 times the original
cloud size. This implies that the momentum had been reduced by a
factor of 36. When the beams' foci were offset, cloud sizes as
much as 10 times the initial cloud size were achieved. In this
later case we observed the atomic cloud to collapse in the $z$
direction when released from certain positions in the trap. An
energy transfer from the $z$ to the $x$ direction was observed for
this release condition for clouds with high densities.

Considering the fact that the cloud's center of mass gains a $z$
velocity this system could be a useful tool in the study of
quantum reflection of cold atoms from material surfaces
\cite{Anderson,Berkhot,Shimizu}. Another possible avenue for
further investigations would be subjecting a Bose-Einstein
condensate to these potentials.

P. Ahmadi was supported by an Oklahoma EPSCoR Nanonet grant and
B.P. Timmons was supported by a NASA Space Grant Fellowship.

\end{document}